\documentclass[preprintnumbers,a4paper,twocolumn,floats,aps,nofootinbib]{revtex4-1}
\usepackage{graphicx}
\usepackage{amsmath,amssymb,amsfonts}
\usepackage{hyperref}
\usepackage{bbm}
\usepackage{xspace}
\usepackage{verbatim}
\usepackage{bm}
\usepackage{color}
\usepackage{cancel}
\usepackage{ulem}
\usepackage{bbm}
\usepackage{xspace}
\usepackage[capitalise, english]{cleveref}
\usepackage[separate-uncertainty=true]{siunitx}
\usepackage{tikz}

\definecolor{tobycolour}{rgb}{.5,.0,.5}

\definecolor{mkgreen}{rgb}{0.2,.70,.3}

\newcommand{\fermiLAT}{Fermi-LAT\xspace}
\DeclareSIUnit\parsec{pc}

\usetikzlibrary{patterns,decorations.pathreplacing}
\usetikzlibrary{decorations.pathmorphing}
\usetikzlibrary{decorations.markings}
\usetikzlibrary{arrows,shapes}
\usetikzlibrary{shapes.misc}
\tikzset{
    gluon/.style={decorate, draw=black,
        decoration={coil,amplitude=4pt, segment length=4pt,aspect=0.7}} 
}
\tikzset{
    photon/.style={decorate, decoration={snake}},
}

\makeatletter
\providecommand*{\diff}%
	{\@ifnextchar^{\DIfF}{\DIfF^{}}}
\def\DIfF^#1{%
	\mathop{\mathrm{\mathstrut d}}%
		\nolimits^{#1}\gobblespace}
\def\gobblespace{%
	\futurelet\diffarg\opspace}
\def\opspace{%
	\let\DiffSpace\!%
	\ifx\diffarg(%
		\let\DiffSpace\relax
	\else
		\ifx\diffarg[%
			\let\DiffSpace\relax
		\else
  			\ifx\diffarg\{%
				\let\DiffSpace\relax
			\fi\fi\fi\DiffSpace}

\begin{document}

\title{Soft Gamma Rays from Heavy WIMPs}

\preprint{CERN-TH-2016-118, Bonn-TH-2016-04}

\author{Manuel Ernst Krauss$^a$}
\email{mkrauss@th.physik.uni-bonn.de}
\author{Toby Opferkuch$^a$}
\email{toby@th.physik.uni-bonn.de}
\author{Florian Staub$^b$}
\email{florian.staub@cern.ch}
\author{Martin Wolfgang Winkler$^a$}
\email{winkler@th.physik.uni-bonn.de}

\affiliation{$^a$Bethe Center for Theoretical Physics \& Physikalisches Institut der 
Universit\"at Bonn, \\
Nu{\ss}allee 12, 53115 Bonn, Germany}
\affiliation{$^b$Theory Department, CERN, 1211 Geneva 23, Switzerland}

\begin{abstract} 
We propose an explanation of the galactic center gamma ray excess by supersymmetric WIMPs as heavy as 500 GeV. The lightest neutralino annihilates into vector-like leptons or quarks which cascade decay through intermediate Higgs bosons. Due to the long decay chains, the gamma ray spectrum is 	much softer than naively expected and peaks at GeV energies. The model predicts correlated diboson and dijet signatures to be tested at the LHC.
\end{abstract}

\maketitle

\section{Introduction}

An excess in the gamma ray flux originating from near the center of the Milky Way was first observed in the \fermiLAT data by Ref.~\cite{Goodenough:2009gk}. In the same work an explanation in terms of annihilating dark matter was proposed. While subsequent analyses have confirmed the significance of the signal~\cite{Hooper:2010mq,Hooper:2011ti,Abazajian:2012pn,Hooper:2013rwa,Gordon:2013vta,Abazajian:2014fta,Daylan:2014rsa,Zhou:2014lva} the modeling of the diffuse gamma ray emission in the galactic center region remained a source of concern. In Ref.~\cite{Calore:2014xka} a comprehensive attempt to quantify the systematic uncertainties in the gamma background was made. The excess remained robust against variations in the diffuse emission models. While plausible astrophysical interpretations in terms of millisecond pulsars~\cite{Abazajian:2010zy} or cosmic ray bursts~\cite{Carlson:2014cwa,Petrovic:2014uda} have been proposed, dark matter annihilation remains one of the prime explanations of the residual signal -- in particular since its morphology matches the expectation from standard dark matter profiles. In Ref.~\cite{Calore:2014xka} it was realized that the inclusion of systematic errors allows for a harder gamma spectrum compared to the early analyses (see also Ref.~\cite{Linden:2016rcf}). This trend continued with the recently published \fermiLAT analysis~\cite{TheFermi-LAT:2015kwa} which includes additional systematic uncertainties beyond Ref.~\cite{Calore:2014xka} in the modeling of the innermost part of the diffuse emission.
The \fermiLAT study was used in Ref.~\cite{Agrawal:2014oha} to show that ordinary WIMPs with masses of $35-\SI{165}{\GeV}$ can fit the gamma ray excess within uncertainties if they annihilate into bottom quarks\footnote{In Ref.~\cite{Agrawal:2014oha} a preliminary version of the \fermiLAT analysis was used \cite{TheFermi-LAT:2014talk}.}. Additionally, dark matter masses of up to $\SI{300}{\GeV}$ became allowed for annihilation into Higgs bosons and top quarks.
In this letter we wish to show that the range of allowed masses extends further as soon as non-standard final states are considered. We focus on the case, where WIMPs annihilate into intermediate particles which then cascade down to standard model states. As a realization we consider the minimal supersymmetric standard model (MSSM) extended by a singlet and a vector-like 5-plet of SU(5)~\cite{Hall:2015xds,Tang:2015eko,Dutta:2016jqn,Barbieri:2016cnt,Nilles:2016bjl,Hall:2016swn}. In UV derived models the presence of vector-like states with TeV masses can be linked to the solution of the supersymmetric $\mu$-problem~\cite{Cvetic:2015vit,Palti:2016kew,Nilles:2016bjl}. Their existence has also been postulated in order to explain the large mass of the standard model Higgs boson~\cite{Basirnia:2016szw}. Given they come in complete GUT multiplets the attractive feature of gauge coupling unification is preserved. The singlet fermion is identified with dark matter and dominantly annihilates into vector-like leptons (or quarks). While in the MSSM, the correct dark matter relic abundance is only realized in very specific corners of the parameter space, the singlet density easily matches $\Omega h^2 \sim 0.12$. The decay chain of the vector-like leptons and further fragmentation results in a continuum of gamma rays which is much softer than that of standard model final states. This remains the case even for dark matter masses $\sim 500\:\text{GeV}$, where the spectrum looks surprisingly similar to the residual spectrum in the galactic center as observed by \fermiLAT~\cite{TheFermi-LAT:2015kwa}.


\section{The Galactic Center Gamma Ray Excess}
\label{sec:galactic}
In order to test the gamma ray spectrum of the proposed model against the galactic center excess, we develop a $\chi^2$ test based on the results of the recent \fermiLAT analysis. Before commenting on the specifics of the analysis we begin by determining the gamma-ray flux and illustrating the choices made for the dark matter distribution and associated uncertainties. 

The gamma-ray flux arising from self-conjugate dark matter particles annihilating into all possible final states $f$ in the Galactic dark matter halo can be expressed as
\begin{align}\label{eq:fluxnormalisation}
\frac{\diff N}{\diff E} &= \frac{1}{2 m_\chi^2}  J  \sum_f \frac{\langle \sigma v \rangle_f}{4\pi} \frac{\diff N_f}{\diff E}  \,.
\end{align}
Here, $m_\chi^2$ is the dark matter mass, while $\langle \sigma v \rangle_f$ and $\diff N_f/\diff E$ are the velocity averaged annihilation cross-section and the spectra of photons obtained for a given final state $f$ respectively. $J$, which contains both the line-of-sight and solid angle integrations, is defined through
\begin{align}
J &= \int_{\Delta \Omega}\diff \Omega\int_{\rm los} \diff s \,\rho\left(r\right)^2 \equiv \mathcal{J} J_{\rm can}\,,
\end{align}
where $\rho(r)$ is the dark matter halo profile centered around the galactic center, $\Delta \Omega$ is the region of interest which for \fermiLAT is a $15^\circ \times 15^\circ$ box around the galactic center, and $s$ is the line-of-sight. 

$J_{\rm can}$ is the central value of $J$. This quantity corresponds to the central values in the parameters describing the dark matter distribution. Therefore $\mathcal{J}$ parametrizes the uncertainty in the dark matter distribution as a deviation from the canonical profile $J_{\rm can}$. Here we choose the (generalized) Navarro-Frenk-White (NFW) profile \cite{Navarro:1995iw}
\begin{align}\label{eq:NFW}
\rho(r)&= \rho_0 \frac{(r/r_s)^{-\gamma}}{(1+r/r_s)^{3-\gamma}} \,,
\end{align}
with the canonical profile corresponding to a local dark matter density of $\rho(r_{\odot})=\SI{0.4}{\GeV\per\cm^3}$ (where $r_{\odot}=\SI{8.5}{\kilo\parsec}$ is the distance between the Sun and the galactic center), scale radius $r_s = \SI{20}{\kilo\parsec}$ and $\gamma=1.2$. We follow the analyses of Refs.~\cite{Calore:2014xka,Calore:2014nla,Iocco:2011jz,Agrawal:2014oha} which estimate the uncertainties to be $\rho(r_{\odot})=\SI{0.4\pm0.2}{\GeV\per\cm^3}$ and $\gamma=1.2\pm0.1$. This translates into the allowed range of $\mathcal{J} \in [0.14,4.4]$ with $J_{\rm can} = \SI{1.08E23}{\GeV^2\per\cm^5}$.

\begin{figure}[htp]
\begin{center}
\includegraphics[width=0.5\textwidth]{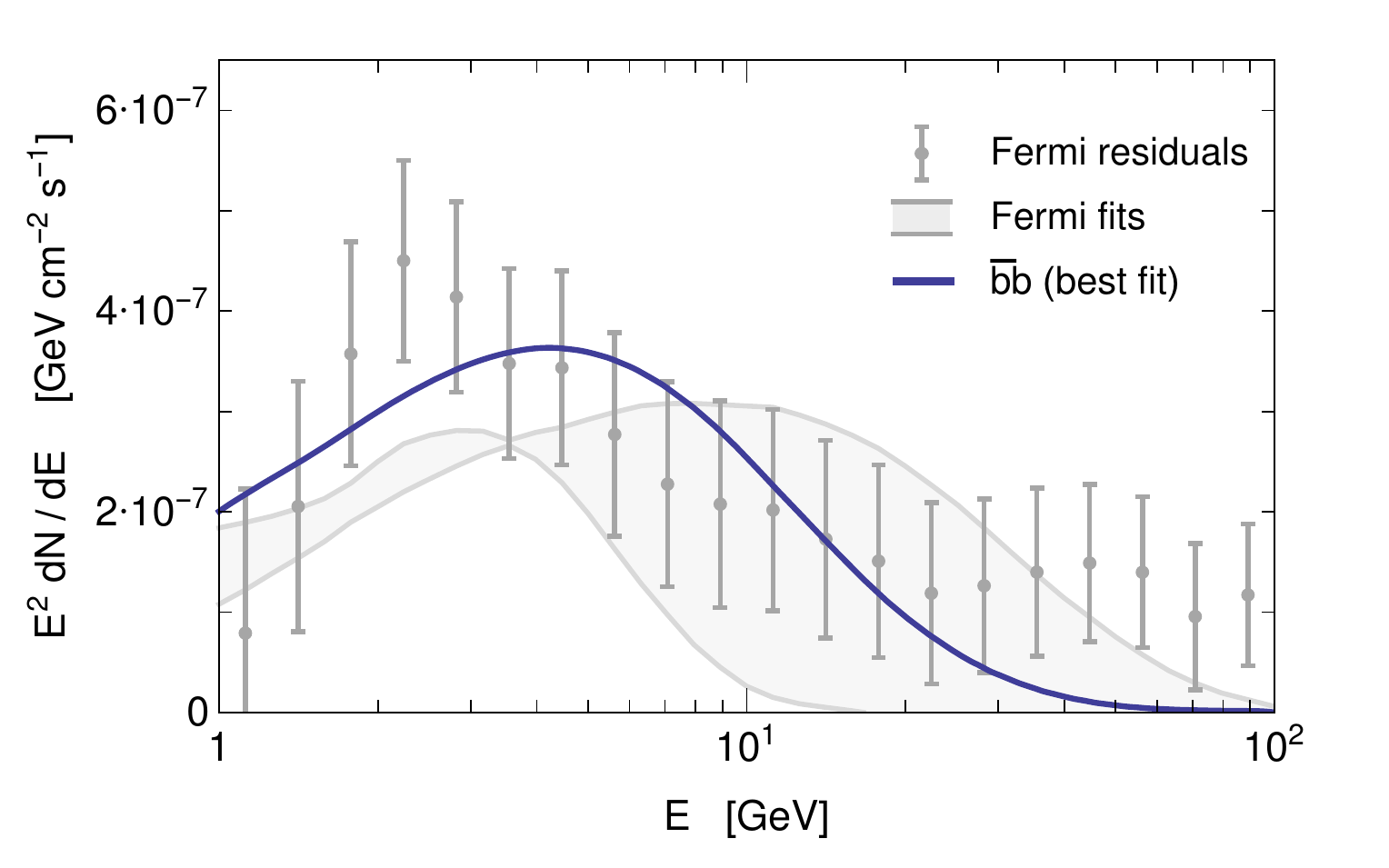}
\end{center}
\caption{Energy spectrum of the galactic center gamma ray excess and uncertainties obtained in our analysis. Also shown is the best fit spectrum for dark matter annihilating into bottom quarks and an envelope showing the spread of the \fermiLAT best-fit exponential cut-off power law spectra.}
\label{fig:bbbarbestfit}
\end{figure}

We now require a goodness-of-fit test for the gamma-ray spectrum based on the results of the \fermiLAT analysis~\cite{TheFermi-LAT:2015kwa}. In this analysis they developed four specialized models for the diffuse gamma-ray background, which when subtracted from the data all lead to non-negligible residuals. We extracted the spectrum of the excess by subtracting the modeled diffuse gamma-ray background, as given in Fig. 12 and 17 of Ref.~\cite{TheFermi-LAT:2015kwa}, from the data. To account for additional systematic uncertainties, we then varied the normalization of the central inverse Compton component by $\pm50\%$ as motivated by the fits in the \fermiLAT analysis \cite{TheFermi-LAT:2015kwa}. Based on the extracted spectra, the mean and variance for each energy bin is determined. The statistical contribution to the variance is estimated from Ref.~\cite{TheFermi-LAT:2014talk} using the number of events per energy bin. In \cref{fig:bbbarbestfit} we depict the final spectrum of the excess including the derived uncertainties. These are then used to define a $\chi^2$ test statistic.

\begin{figure}[htp]
\begin{center}
\includegraphics[width=0.5\textwidth]{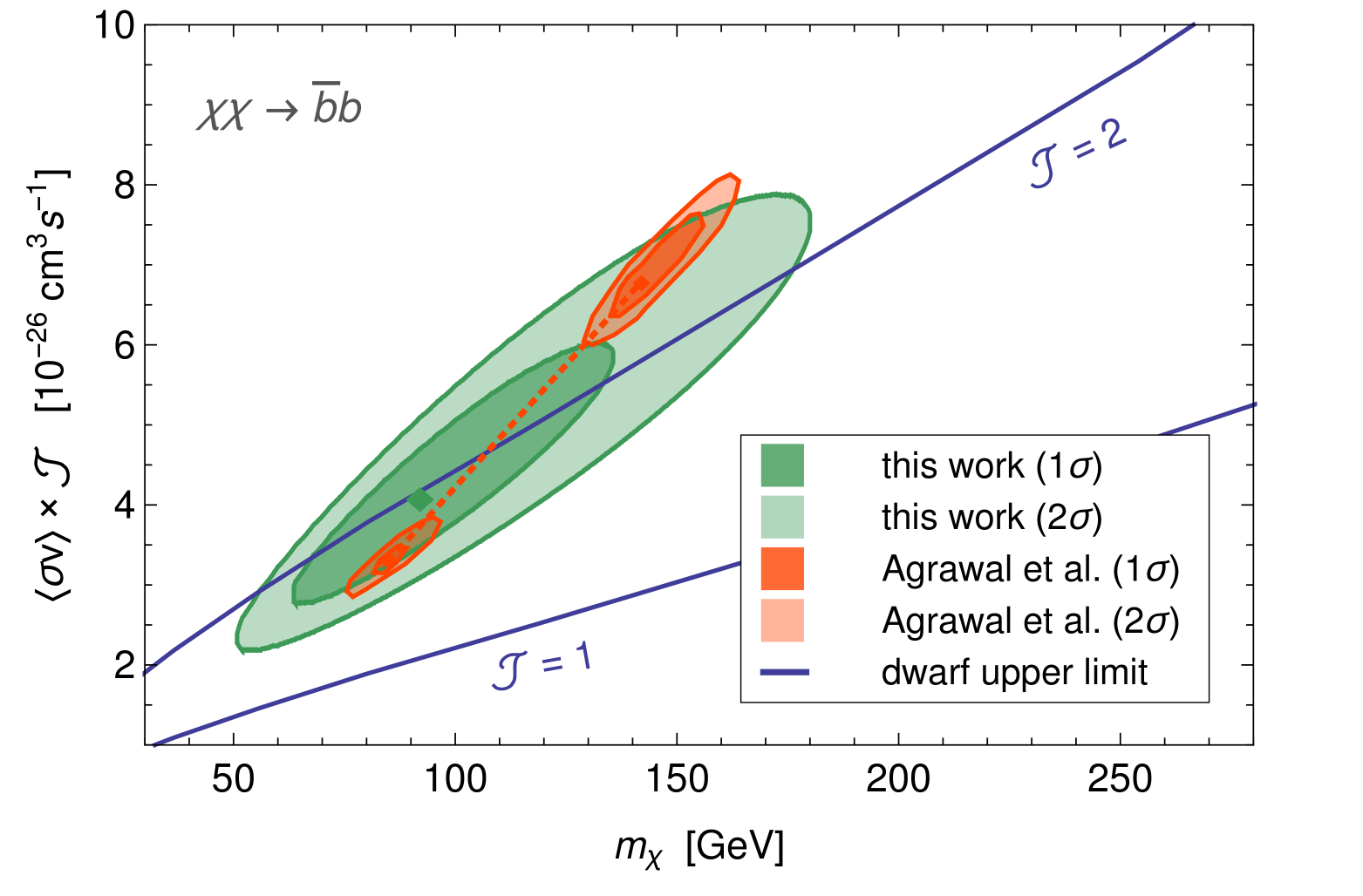}
\end{center}
\caption{Parameter regions consistent with the \fermiLAT gamma-ray excess at $1\,\sigma$ and $2\,\sigma$. Annihilation of dark matter into bottom quarks is assumed. The green regions refer to our analysis, while the orange regions were obtained in Ref.~\cite{Agrawal:2014oha} for two different diffuse emission templates (see text).}
\label{fig:bbbarcontours}
\end{figure}

To validate our interpretation of the \fermiLAT analysis we perform a $\Delta\chi^2$ fit under the assumption that dark matter annihilates purely to a $\bar{b}b$ final state. This final state is not directly related to our scenario and only serves for the comparison with previous analyses.
The resulting gamma-ray spectrum as a function of the dark matter mass is taken from Refs.~\cite{Cirelli:2010xx,Ciafaloni:2010ti}. Utilizing the flux normalization given in \cref{eq:fluxnormalisation}, the parameter regions consistent with a $\Delta \chi^2$ at the 1 and 2$\sigma$ level as well as the best fit point, with $\chi^2_{\rm min}/\text{d.o.f.}=0.98$, are shown in \cref{fig:bbbarcontours}. Also shown in this figure are the resulting best fit contours from Ref.~\cite{Agrawal:2014oha}. Here the two disconnected contour regions are the result of fitting two different best fit spectra from the \fermiLAT analysis. The authors argue that the region connecting these two contours is also allowed as each contour represents an extreme choice of the interstellar emission model. By including all diffuse interstellar emission models and varying the inverse Compton component our analysis effectively encompasses uncertainties arising through this choice of model. Also included in \cref{fig:bbbarcontours} are the limits on the annihilation cross section from dwarf galaxies. For the canonical dark matter profile ($\mathcal{J}=1$) these exclude the entire parameter space favored by the galactic center excess. However, a slightly steeper profile ($\mathcal{J} \gtrsim 2$) -- well within uncertainties -- reconciles the excess with the dwarf limits. The spectrum of the best-fit $\bar{b}b$ point (as indicated in \cref{fig:bbbarcontours}) is depicted as a blue line in \cref{fig:bbbarbestfit}. In addition we show our derived energy spectrum of the galactic center excess along with an envelope which encompasses the different \fermiLAT spectra assuming an exponential cut-off power law. These were obtained in Ref.~\cite{TheFermi-LAT:2015kwa} by fitting an excess component with this spectral form to the four mentioned diffuse emission models.
While these fits have insufficient spectral freedom to pick up all residuals, it is noted in Ref.~\cite{TheFermi-LAT:2015kwa} that they cannot be ruled out due to the limitations in modeling the interstellar emission. In our goodness-of-fit test they would receive $\chi^2/\text{d.o.f.}$ between 1.1 and 2.1.

\section{The Galactic Center Excess in a Supersymmetric Model}
\label{sec:model}

We consider the MSSM extended by a singlet as well as one copy of vector-like matter superfields transforming as ${\bf 5} + \bf{\bar{5}}$ under $SU(5)$. These contain a
lepton-like $SU(2)_L$ doublet, $\hat L_5 = (\hat \nu_5,\hat \ell_5)$, as well as 
a right-handed down-quark-like superfield, $\hat D_5$. The corresponding superpotential reads
\begin{align}
W &= W_{\rm MSSM} + \frac{\mu_S}{2} \hat S^2 + \kappa \hat S^3 + (\mu_L + \lambda_L \hat S) \hat{\bar{L}}_5 \hat L_5 \label{eq:superpotential}\\ 
\notag &+ (\mu_D + \lambda_D \hat S) \hat{\bar{D}}_5 \hat D_5 + Y_D' \hat D_5 \hat Q \hat H_d + Y_L' \hat E \hat L_5 \hat H_d\,, 
\end{align}
where we have suppressed color, $SU(2)$ and generation indices.
Here, $W_{\rm MSSM}$ denotes the MSSM superpotential and $\hat Q$ and $\hat E$ the MSSM quark doublet and right-handed charged lepton superfields, respectively. In our notation, hats are assigned to all superfields; in what follows, $R$-parity-odd fields receive a tilde. The presence of the $Y'$ couplings avoids a $Z_2$ symmetry and allows the decay of the otherwise stable vector-like states. In general, $Y'$ has three entries, but we make the assumption that only the third component is non-vanishing to avoid limits from flavor changing observables. The scalar singlet can be decomposed into its CP-even and CP-odd components $h_s$ and $a_s$. In the absence of a $\hat S \hat H_u \hat H_d$ term in the superpotential, mixing of the singlet with the MSSM Higgs bosons can be neglected. For our analysis we have used the {\tt Mathematica} package {\tt SARAH} \cite{Staub:2008uz,Staub:2009bi,Staub:2010jh,Staub:2012pb,Staub:2013tta,Staub:2015kfa} and slightly modified the {\tt NMSSM+VL/5plets} model as provided with Ref.~\cite{Staub:2016dxq}. The interface to the spectrum generator {\tt SPheno} \cite{Porod:2003um,Porod:2011nf} enables the computation of the mass spectrum \cite{Goodsell:2014bna,Goodsell:2015ira}.

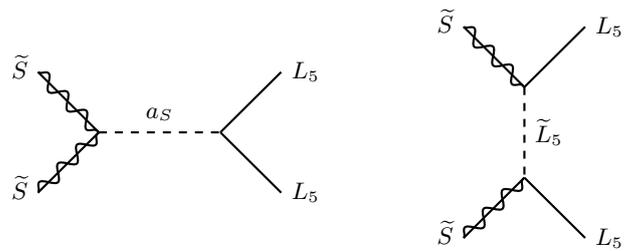
\begin{figure}
\centering
\begin{tikzpicture}
[scale=1.6]
\begin{scope}[thick,decoration={
    markings,
    mark=at position 0.8 with {\arrow{stealth}}}
    ]

\draw[-] (0,1)--(0.5,0.5); 
\draw[-] (0,0)--(0.5,0.5); 
\draw[photon] (0,1)--(0.5,0.5); 
\draw[photon] (0,0)--(0.5,0.5); 
\draw[-,dashed] (0.5,0.5)--(1.5,0.5); 
\draw[-] (2,1)--(1.5,0.5); 
\draw[-] (2,0)--(1.5,0.5); 

\node[black] at (-0.15,0.04) {$\widetilde{S}$};
\node[black] at (-0.15,1.04) {$\widetilde{S}$};
\node[black] at (1,0.65) {$a_S$};
\node[black] at (2.2,0.00) {$L_5$};
\node[black] at (2.2,1.00) {$L_5$};

\draw[-,dashed] (4.0,1.25-0.375)--(4.0,0.5-0.375); 
 
\draw[photon] (3.5,0-0.375)--(4.0,0.5-0.375); 
\draw[-] (3.5,0-0.375)--(4,0.5-0.375); 
\draw[-] (4,0.5-0.375)--(4.5,0-0.375); 

\draw[photon] (3.5,1.75-0.375)--(4,1.25-0.375); 
\draw[-] (3.5,1.75-0.375)--(4,1.25-0.375); 
\draw[-] (4,1.25-0.375)--(4.5,1.75-0.375); 

\node[black] at (3.35,0-0.375+0.05) {$\widetilde{S}$};
\node[black] at (3.35,1.75-0.375+0.05) {$\widetilde{S}$};
\node[black] at (4.2,0.5) {$\widetilde{L}_5$};
\node[black] at (4.70,1.75-0.375) {$L_5$};
\node[black] at (4.70,0-0.375) {$L_5$};
\end{scope}

\end{tikzpicture}
\caption{Relevant diagrams for singlino dark matter annihilation.}
\label{fig:dm_annihilation_diagrams}
\end{figure}

We consider the scenario where the singlet fermion, $\tilde S$, is the dark matter candidate. The dominant annihilation final states, if kinematically allowed, are pairs of vector-like leptons or quarks which then further decay via the couplings $Y_L'$ and $Y_D'$. Due to the strong LHC constraints on vector-like quark masses, we will focus on the case where only the vector-like leptons are lighter than $\tilde S$. The annihilation then proceeds via $t$-channel $\tilde L_5$ exchange and $s$-channel mediation via the pseudoscalar $a_s$ (see Fig.~\ref{fig:dm_annihilation_diagrams}). For $t$-channel exchange, $\lambda_L$ and the mass of the slepton determine the efficiency of the annihilation, while $s$-channel mediation requires a non-negligible self-coupling $\kappa$ for the $\tilde S\, \tilde S\, a_s$ vertex. For the calculation of the dark matter relic density and the gamma ray spectrum, we use  {\tt MicrOMEGAs-4.2.5} \cite{Belanger:2001fz,Belanger:2004yn,Belanger:2006is,Belanger:2008sj,Belanger:2013oya,Belanger:2014vza} through the interface with {\tt SARAH}. As the annihilation does not involve highly energetic electrons or muons, we neglect bremsstrahlung and inverse Compton contributions.\footnote{The latter could become important if $Y_L'$ contained non-vanishing first or second entries opening the decay of $L_5$ into electrons or muons.} The dark matter distribution is modeled with the NFW profile as described in Eq.~\ref{eq:NFW}. The goodness-of-fit to the galactic center excess is obtained from our $\chi^2$ measure as described in Section~\ref{sec:galactic}.

The decay pattern of the vector-like leptons strongly depends on kinematics. In Tab.~\ref{tab:leptondecay} we specify the branching ratios for three different masses. As the vector-like leptons couple to $H_d$, decays into the heavy CP even, CP odd and charged MSSM Higgs bosons $H,\,A$ and $H^\pm$ dominate as long as they are kinematically accessible. Below their production threshold, the relevant modes involve the light Higgs or an electroweak gauge boson. For $m_{L_5} < 125 \:\text{GeV}$ only the decays into gauge bosons survive.

\begin{table}[ht]
\begin{center}
\begin{tabular}{cccc}
\toprule
$m_{L_5}$ & $105\:\text{GeV}$ & $250\:\text{GeV}$ & $450\:\text{GeV}$\\
\colrule
$\ell_5 \to \tau + Z$  & $99.8\%$  & $54.7\%$ & $12.5\%$   \\ 
$\ell_5 \to \tau + h$  & $-$  & $45.3\%$ & $15.0\%$   \\ 
$\ell_5 \to \tau + H$  & $-$  & $-$ & $35.2\%$   \\ 
$\ell_5 \to \tau +A$  & $-$  & $-$ & $37.3\%$  \\  \colrule
$\nu_5 \to \tau + H^\pm$  & $-$  & $-$ & $73.3\%$ \\
$\nu_5 \to \tau + W$  & $100\%$  & $100\%$ & $26.7\%$ \\ \botrule
\end{tabular}
\end{center}
\caption{Branching ratios of the vector-like leptons for three different masses. The heavy MSSM Higgs boson masses are taken to be $m_A=360$~GeV and $m_H=367$~GeV.}
\label{tab:leptondecay}
\end{table}

\begin{figure}
\includegraphics[width=\linewidth]{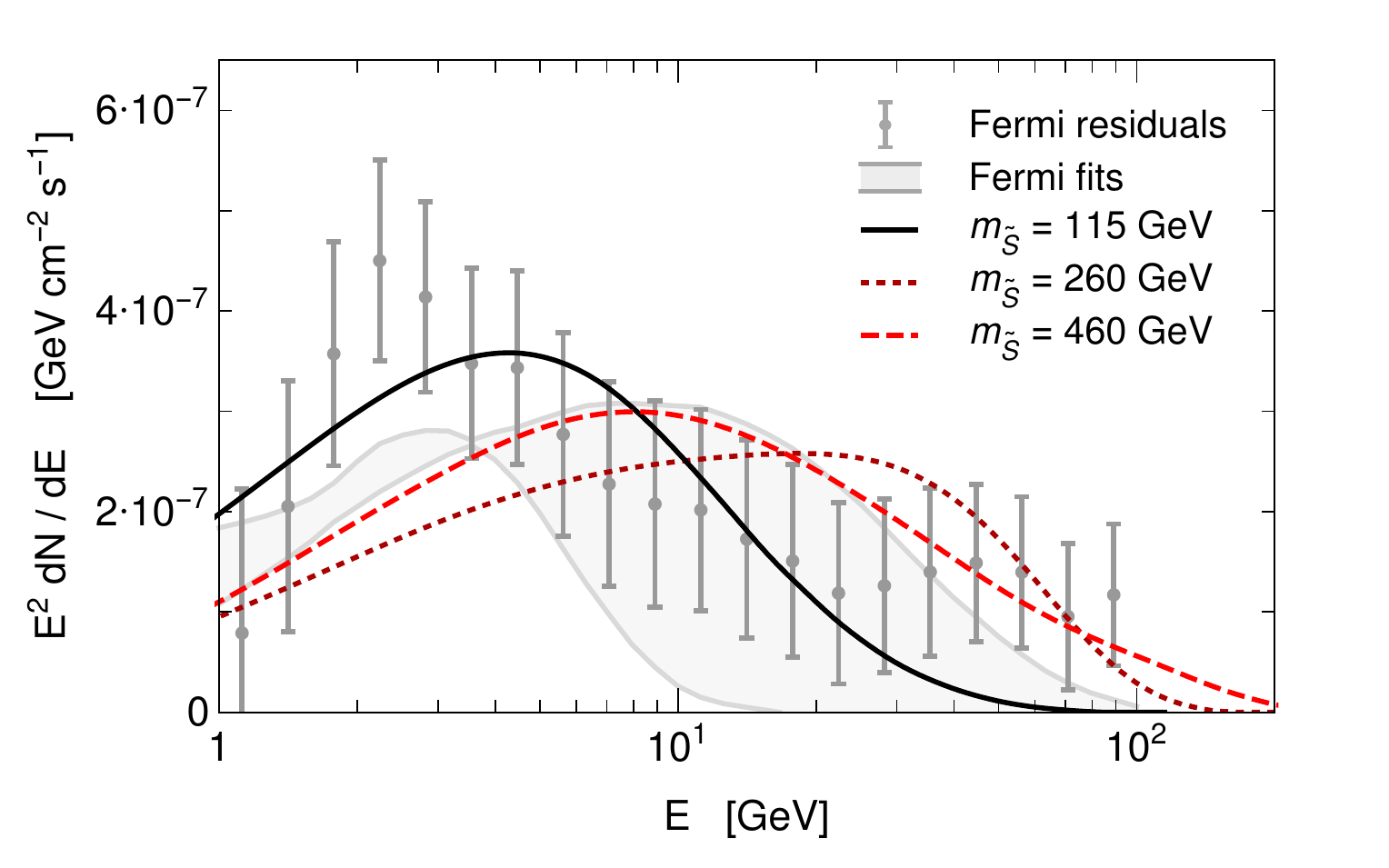}
\caption{Energy spectrum of the galactic center gamma rays for three different singlino masses.
Also shown are the residuals as obtained in Section~\ref{sec:galactic}
as well as the envelope of the \fermiLAT
best-fit spectra assuming an exponential cut-off power law spectrum.}
\label{fig:spectra_resonance}
\end{figure}

The MSSM Higgs bosons dominantly decay as $h \to b \bar b$, $H^\pm \to t b$ and $H/A \to t \bar{t} \text{ or } b \bar b$ depending on kinematics. It is mainly the combination of the masses $m_{\tilde S}$, $m_{L_5}$ and $m_{A}$ which fixes the gamma ray spectrum. In order to match the \fermiLAT excess the gamma ray flux should be rather soft, peaking at GeV energies, see Fig.~\ref{fig:bbbarbestfit}.  This is a non-trivial constraint if we consider dark masses up to several hundreds of GeV. First of all, a large mass gap between the singlino and the vector-like leptons should be avoided as it would lead to undesirable smearing of the spectrum. Even if the dark matter mass resides close to $m_{L_5}$, the vector-like lepton decay leads to a strong energy release. Fortunately, through the following hadronic cascade this energy is distributed between a large number of final states. In Fig.~\ref{fig:spectra_resonance} we depict the gamma ray flux for the three example spectra of Tab.~\ref{tab:leptondecay} where we set $m_{\tilde S} = m_{L_5}  + 10\:\text{GeV}$. The normalization factor $\mathcal J$ of each spectrum is chosen to minimize $\chi^2$. The light singlino case provides a good fit $\chi^2/\text{d.o.f.} =0.93$ to the \fermiLAT excess with the gamma rays mainly stemming from electroweak gauge bosons. Indeed, the spectrum resembles the one of dark matter directly annihilating into $WW$ bosons (see e.g.\ \cite{Agrawal:2014oha}). For the intermediate singlino mass, the fit becomes poor ($\chi^2/\text{d.o.f.} =1.57$) as the decay products of the vector-like leptons experience a significant boost. The shoulder at the end of the spectrum origins from the highly energetic taus. For even higher singlino masses the fit improves again to $\chi^2/\text{d.o.f.} =1.02$. For this mass the decays of the vector-like leptons into heavy Higgs bosons become kinematically accessible. Consequently the number of final states in the hadronic showers increases and the gamma ray spectrum becomes softer again.

\begin{figure}[ht]
\begin{center}
\includegraphics[width=0.8\linewidth]{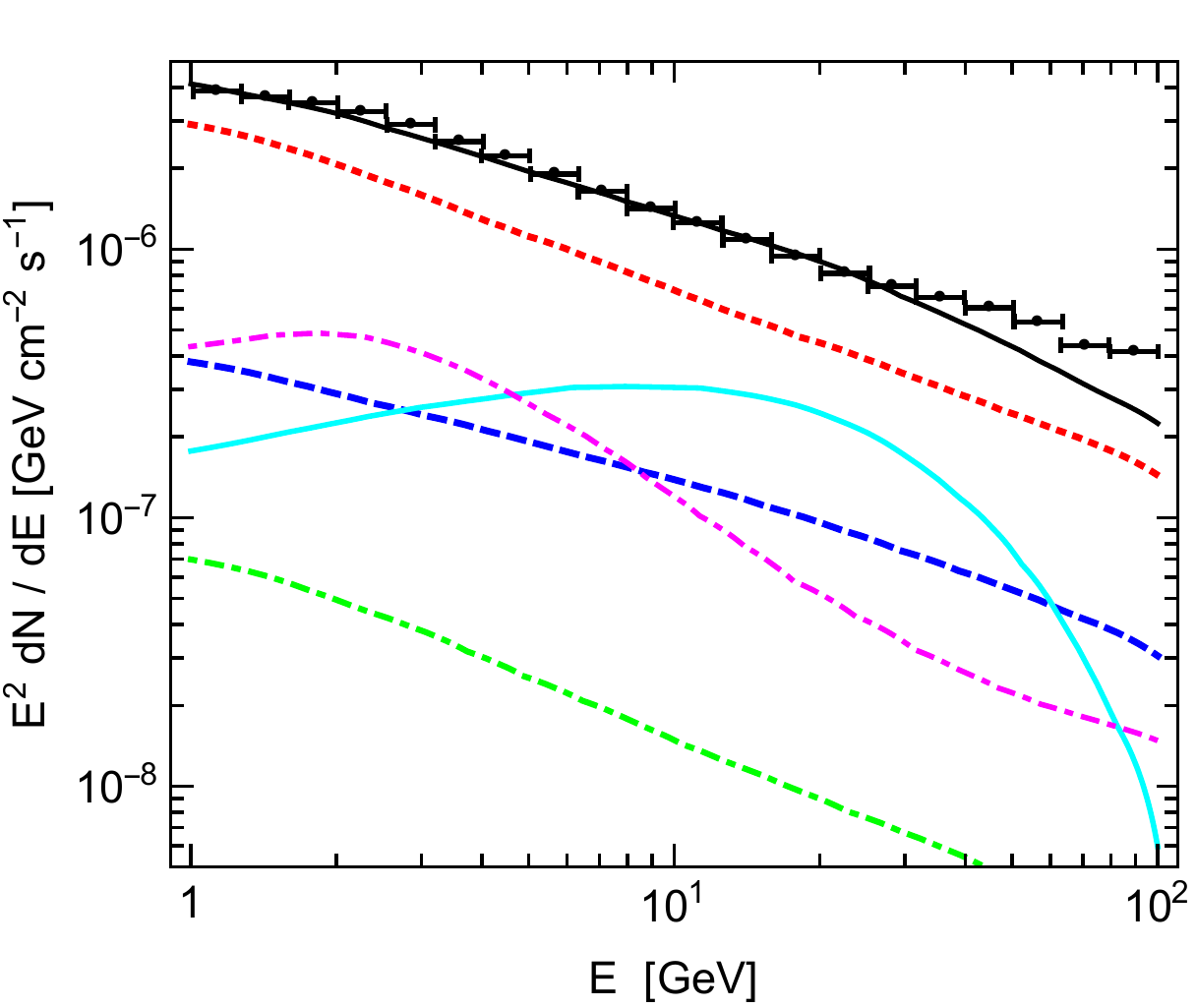}\\
\includegraphics[width=0.8\linewidth]{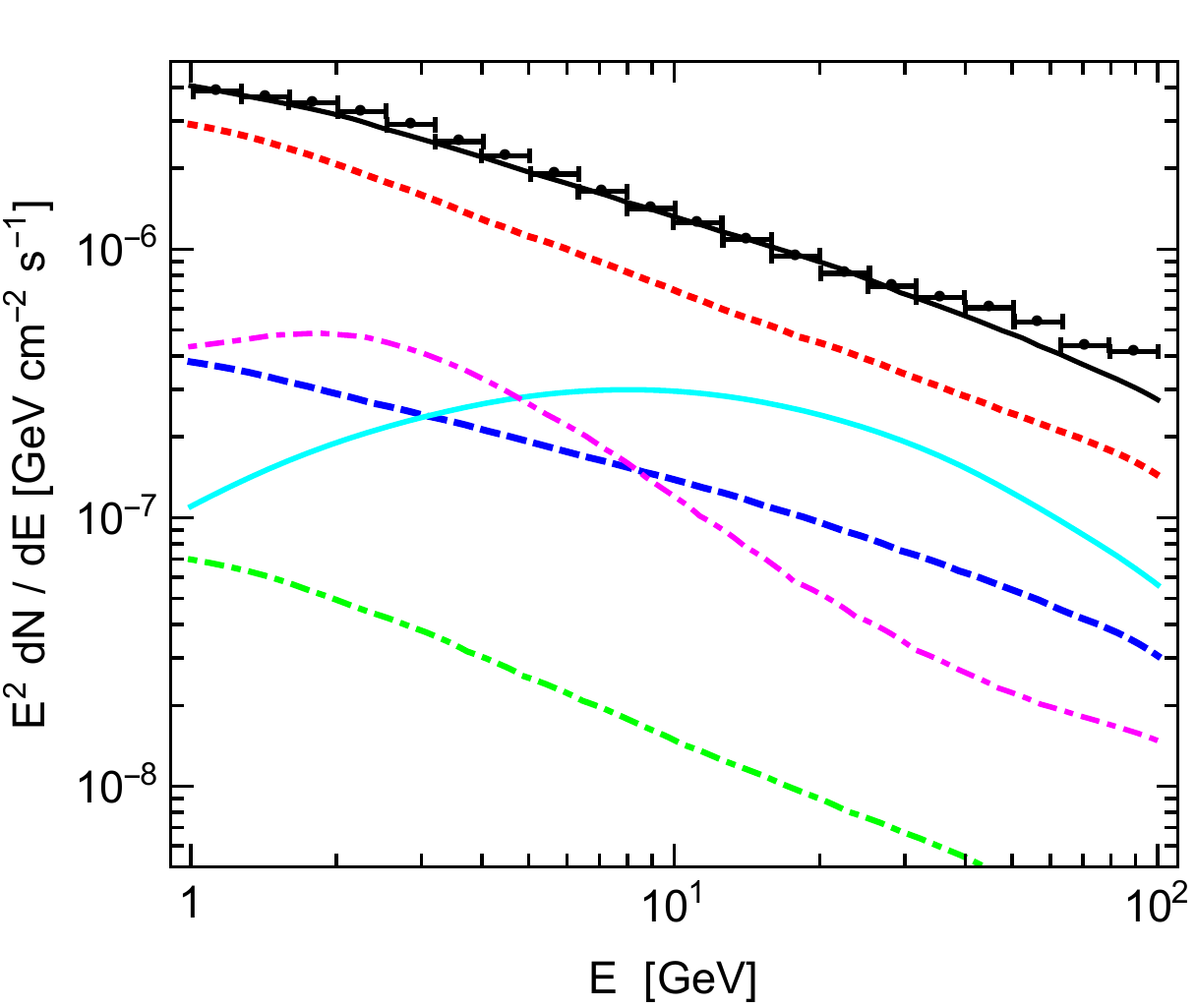}\\[3mm]
\hspace{1cm}\includegraphics[width=0.45\linewidth]{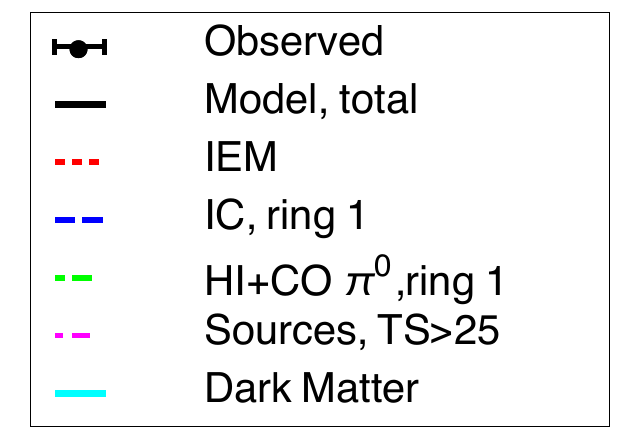}
\end{center}
\caption{Galactic center gamma ray emission including a dark matter component (cyan line) for the \fermiLAT intensity scaled pulsar model of interstellar emission (see~\cite{TheFermi-LAT:2015kwa} for details). In the upper panel the dark matter signal refers to the best-fit exponential cut-off power law spectrum, in the lower panel to the predicted spectrum in our model (heavy singlino case).}
\label{fig:fermi_data_comparison}
\end{figure}

While the spectrum with the largest singlino mass falls into the $1\,\sigma$ range of our goodness-of-fit test it appears not to capture the residual spectrum at $E\lesssim 2\:\text{GeV}$ very well. However, it should be noted that gamma ray spectra from dark matter annihilation rather generically take the (approximate) form of an exponential cut-off power law~\cite{Agrawal:2014oha}. As mentioned, \fermiLAT has provided best fit spectra of this form for their different diffuse background models, their resulting envelope is also shown in Fig.~\ref{fig:spectra_resonance}. Our high mass spectrum bears resemblance with the best exponential cut-off power law spectrum for the intensity scaled pulsar model of interstellar emission as we show in Fig.~\ref{fig:fermi_data_comparison}. The dark matter spectrum with the light singlino fits the low energy part of the excess better, but the trade-off is a worse fit at higher energies (similar as for annihilation into bottom quarks, see Fig.~\ref{fig:bbbarcontours}). Hence, we regard the heavy singlino as an equally valid explanation of the galactic center excess.

We shall now turn to the normalization of the gamma ray spectrum and apply the constraints from dwarf galaxies.\footnote{The dwarf limits are only available for limited types of dark matter annihilation final states like $b \bar b$ and $W^+W^-$. As the latter spectrum is most similar in shape to our scenario, we infer the corresponding bounds by fitting our spectrum to a $WW$ spectrum with a free dark matter mass and annihilation cross section.} To make the singlino a thermal dark matter candidate its freeze-out cross section must be $\langle \sigma v\rangle_{\rm FO} = 2-3 \times 10^{-26} \:\text{cm}^3/\text{s}$. This can easily be achieved through the diagrams of Fig.~\ref{fig:dm_annihilation_diagrams} with an appropriate choice of the coupling $\lambda_L$.  Given today's annihilation cross section matches the one at freeze-out, the normalization of the gamma ray flux is fixed up to the uncertainty $\mathcal{J}$ in the dark matter profile. For the light singlino, the best fit gamma ray spectrum of Fig.~\ref{fig:spectra_resonance} is obtained for $\mathcal{J}=3.0$ if a thermal cross section is imposed. This value is large enough to satisfy the constraints from dwarf galaxies and, at the same time, is well within the astrophysical uncertainties $\mathcal{J} \in [0.14,4.4]$ we consider as realistic. Hence, the light singlino provides an attractive explanation to the galactic center excess in terms of a thermal WIMP.

As the intermediate mass singlino is not of interest due to its poor fit to the \fermiLAT spectrum, we directly turn to the heavy singlino case. Given a thermal cross section, the best fit gamma ray spectrum of Fig.~\ref{fig:spectra_resonance} requires $\mathcal{J}=8.8$. This is somewhat beyond the allowed range of $\mathcal{J}$. We checked that the problem persists for all spectra with heavy singlinos $m_{\tilde{S}} > 300\:\text{GeV}$. In this mass window -- if one is not willing to push astrophysical uncertainties, e.g.\ by invoking more strongly contracted dark matter profiles -- one needs to resort to mechanisms that enhance today's annihilation cross section over the one at freeze-out. In the following we discuss two explicit mechanisms which result in $\Omega h^2$ within the measured range \cite{Ade:2013zuv} as well as a proper normalization to the galactic center excess with $\mathcal{J}\lesssim 4$: bino coannihilation and a pseudoscalar resonance.

\begin{figure}[t]
\includegraphics[width=\linewidth]{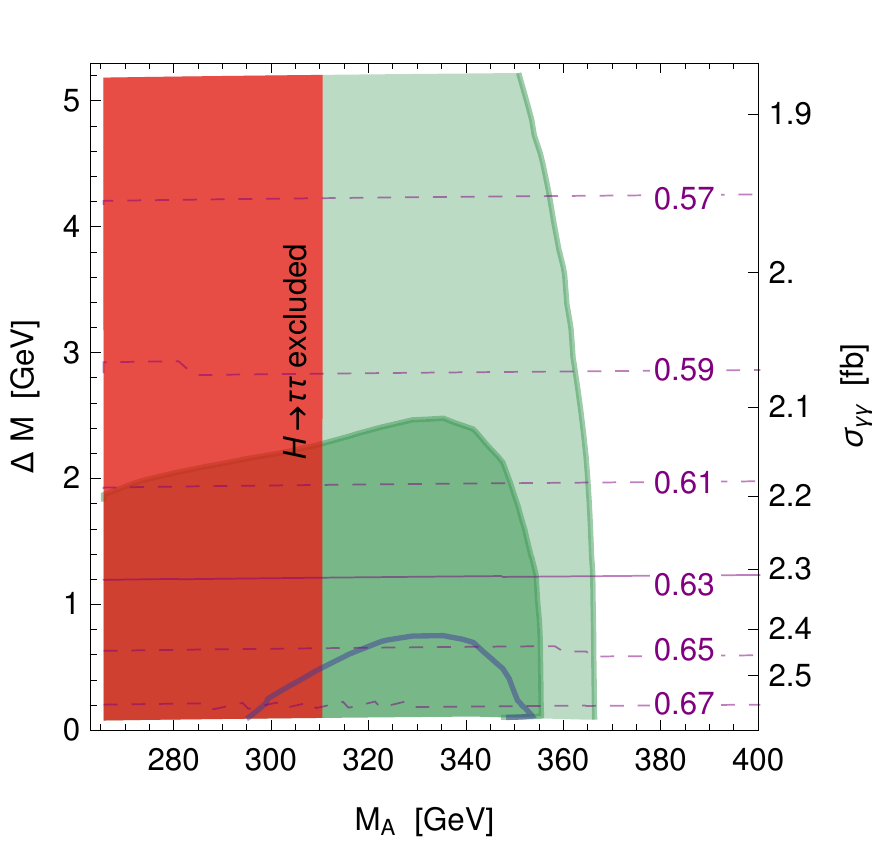}
\caption{
Parameter scan indicating gamma ray spectra which fit the galactic center excess at
$1\,\sigma$ (dark green) and $2\,\sigma$ (light green) depending on the MSSM pseudoscalar mass and the bino-singlino mass splitting. We set $m_{\tilde S }= 387~{\rm GeV},\,m_{L_5}=382~{\rm GeV},\,m_{\tilde L_{5;1,2}}\approx 500~{\rm GeV},\,\tan \beta=7.5,\,\kappa=0$. The relic density is fixed at $\Omega h^2\approx 0.12$ by adjusting the coupling $\lambda_L$ as indicated by the purple contours. On the right $y$-axis we show the maximal diphoton cross section at 13~TeV from the added signals of $h_s$ and $a_s$, see Sec.~\ref{sec:collider}. The area marked in red is in conflict with LHC searches for a heavy Higgs decaying to $\tau \tau$ \cite{CMS:2015mca}. The blue contour at the bottom would be obtained if the current gamma ray limits from dwarf galaxies became stronger by 50\%. }
\label{fig:bino_coannihilation}
\end{figure}

\begin{figure}
\includegraphics[width=\linewidth,trim={1.4cm 0 1.4cm 0},clip]{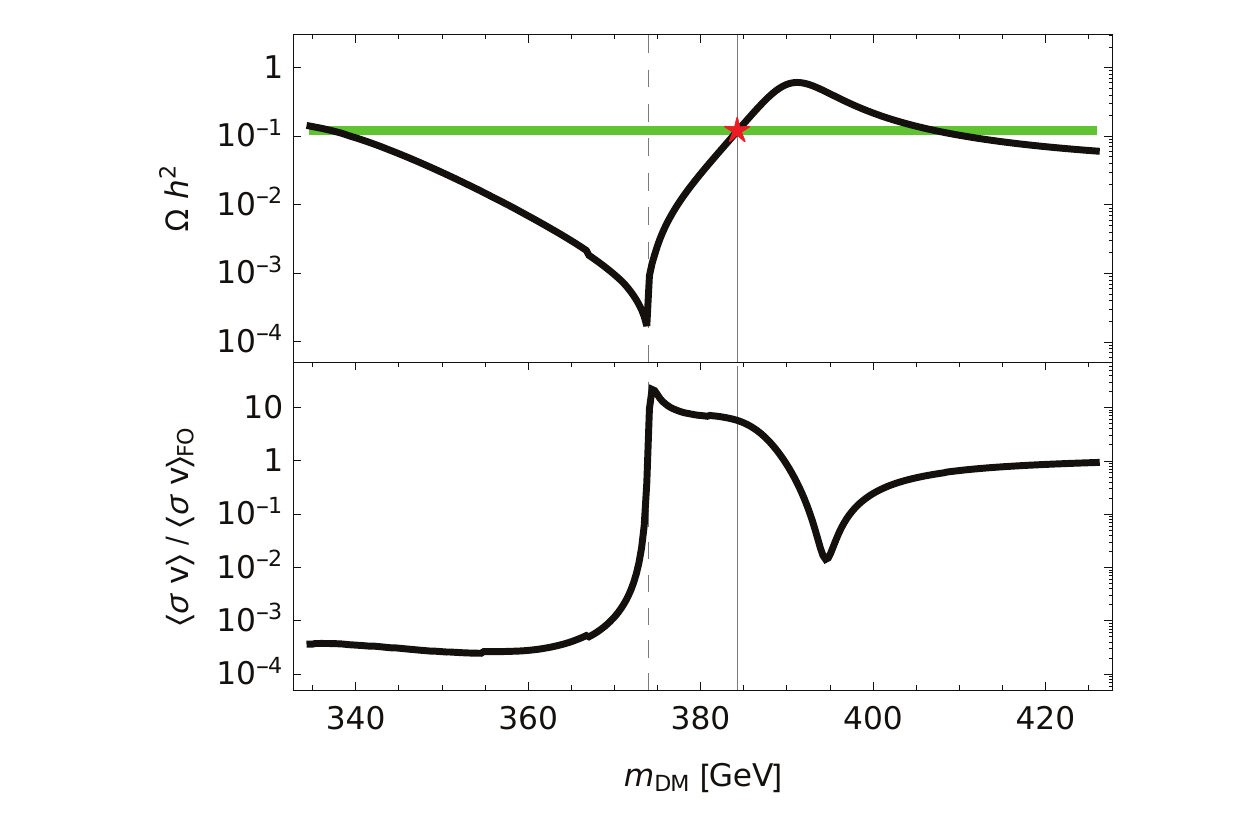}
\caption{Dependence of the relic density and the ratio of velocity-averaged annihilation cross sections today and at freeze-out on the mass of the singlino with fixed $m_{a_s}=748~{\rm GeV}$ and $m_{L_5}=374.5~$GeV, using $\kappa=2.4 \cdot 10^{-2}$, $m_A=334~{\rm GeV},\,\tan \beta = 7,\, \lambda_L=0.63$. The green line indicates the observed dark matter relic density.}
\label{fig:resonance750}
\end{figure}

\paragraph*{Bino coannihilation}  For a small mass splitting $\Delta M$ between $\tilde S$ and the bino $\tilde B$, coannihilation becomes important for the thermal cross section. As the bino interacts weaker than the singlino, this will in total reduce $\langle \sigma v\rangle_{\rm FO}$. If one then adjusts the coupling $\lambda_L$ to keep the relic density fixed at $\Omega h^2\approx 0.12$, one effectively enhances today's annihilation $\langle \sigma v\rangle$ compared to the case without coannihilations. For the heavy singlino spectrum of Fig.~\ref{fig:spectra_resonance} this mechanism allows to reduce the required $\mathcal{J}$ down to $\mathcal{J}=3.1$. For further illustration, Fig.~\ref{fig:bino_coannihilation} shows a scan over $m_A$ and $\Delta M$ while keeping the singlino mass and the relic density fixed through adapting $\lambda_L$. Requiring $\mathcal{J}\leq 4.4$ the galactic center excess is fit within $1\,\sigma$ for mass splittings up to $\Delta M \sim 2.5\:\text{GeV}$ and not too large $m_A$. The region at $m_A\gtrsim m_{\tilde{S}}$ is excluded as the gamma spectrum becomes too hard. In this range the decay of the vector-like leptons via $A,\, H$ is phase-space suppressed or forbidden and the decay products are very boosted. Small $m_A\lesssim 300 \:\text{GeV}$ is, furthermore, excluded by LHC Higgs searches in the $\tau\tau$-channel~\cite{CMS:2015mca}.\footnote{We chose a value of $\tan\beta=7$ and obtain the limit with the tool {\tt HiggsBounds} \cite{Bechtle:2008jh,Bechtle:2011sb,Bechtle:2013wla}.} Another constraint $\Delta M \gtrsim 1\:\text{GeV}$ arises if one requires the model to remain perturbative up to the GUT scale. At smaller $\Delta M$, the correct relic density enforces a value of $\lambda_L>0.63$ which would lead to a Landau pole in the RGE running of this coupling at intermediate scales~\cite{Hall:2015xds,Nilles:2016bjl}. Gamma ray limits from dwarf galaxies are close in sensitivity but do not yet lead to an exclusion. Hence, it survives a viable region of parameter space in which a thermally produced heavy singlino can explain the galactic center excess while satisfying all constraints.\\ 
\paragraph*{Pseudoscalar resonance}  A further possibility to enhance $\langle \sigma v\rangle/\langle \sigma v\rangle_{\rm FO}$ opens up near the pseudoscalar $s$-channel resonance if the mass splitting $m_{a_s}-2 m_{L_5}$ is sufficiently small: while today's annihilation cross section $\tilde S \tilde S \to a_s$ peaks at $m_{\tilde S} = m_{a_s}/2$, the thermal cross section already peaks at slightly smaller dark matter masses due to the higher momentum in the early universe. Because of this relative shift of the peaks, today's annihilation cross section is enhanced with respect to the one at freeze-out for $m_{\tilde S} \gtrsim m_{a_s}/2$. In Fig.~\ref{fig:resonance750} we scan the singlino mass around $m_{a_s}/2$.
We show the relic density as well as $\langle \sigma v\rangle/\langle \sigma v\rangle_{\rm FO}$ as a function of the singlino mass using $\kappa = 2.4 \cdot 10^{-2}$. The apparent broad width of the resonance at lower masses is due to the open decay channels $a_s \to \tilde S \tilde S$. Above the resonance, where $\langle \sigma v\rangle/\langle \sigma v\rangle_{\rm FO}$ reaches its maximal value, the $a_s$ decay into two singlinos is forbidden but $a_s \to L_5 \bar L_5$ becomes kinematically allowed. A parameter point with the correct dark matter density is marked by the solid gray line, indicating a relative cross-section enhancement of  $\langle \sigma v\rangle/\langle \sigma v\rangle_{\rm FO} \approx 5$. The corresponding gamma spectrum fits the galactic center excess within $1\,\sigma$. For singlino masses around 390~GeV, a destructive interference between the $s$- and $t$-channel diagrams leads to a small peak (dip) in $\Omega h^2$ (the cross section ratio). At higher dark matter masses, the annihilation mechanism is dominated by the $\tilde L_5$ $t$-channel and $\langle \sigma v\rangle/\langle \sigma v\rangle_{\rm FO}$ approaches unity.

Finally, we have checked that the monochromatic diphoton line which is emitted by singlino annihilation at loop order is consistent with existing bounds \cite{Ackermann:2015lka} for all considered spectra. In addition, we have used {\tt MicrOMEGAs} to calculate the dark matter induced antiproton flux taking propagation parameters from the recent boron to carbon analysis~\cite{Kappl:2015bqa}. The predicted flux is not excluded but close to the sensitivity of AMS-02~\cite{AMS:2015talk}.

\section{Collider Signatures}
\label{sec:collider}
In this section we consider collider constraints and possible new signals of the particles introduced beyond the MSSM. The most stringent constraints in this new sector arise on vector-like quark masses. These limits require $m_{D_5}\gtrsim 750\:\text{GeV}$~\cite{Aad:2015kqa}. In contrast, there are no direct LHC constraints on heavy charged vector-like leptons. The best constraint originates from LEP searches for heavy charged leptons \cite{Achard:2001qw} which imposes the bound $m_{L_5} \gtrsim  100\:\text{GeV}$. 

The scalars $h_s$ and $a_s$ can be produced at the LHC through loops of both the fermionic and scalar components of the vector-like quark superfields. The typical signature depends on the mass of the vector-like leptons. If decays of $h_s$ and $a_s$ to the vector-like leptons are kinematically allowed then the final states are determined by the branching ratios shown in Tab.~\ref{tab:leptondecay}. For instance if $m_{L_5}=105$~GeV then the two dominant final states would be $\tau^+\tau^-ZZ$ and $\nu_\tau \bar{\nu}_\tau W^+ W^-$ for the charged and neutral vector-like lepton mediators respectively. However, both of these final states are not directly constrained by current LHC searches.   

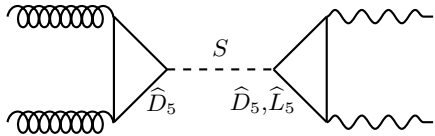
\begin{figure}
\centering
\begin{tikzpicture}
\begin{scope}[thick]

\draw[gluon] (0,1*1.4)--(1*1.4,1*1.4);
\draw[gluon] (0,0)--(1*1.4,0);

\draw[-] (1*1.4,0)--(1*1.4,1*1.4);
\draw[-] (1*1.4,0)--(1.5*1.4,0.5*1.4);
\draw[-] (1*1.4,1*1.4)--(1.5*1.4,0.5*1.4);
\draw[-,dashed] (1.5*1.4,0.5*1.4)--(2.5*1.4,0.5*1.4);
\draw[-] (2.5*1.4,0.5*1.4)--(3*1.4,1*1.4);
\draw[-] (2.5*1.4,0.5*1.4)--(3*1.4,0);
\draw[-] (3*1.4,1*1.4)--(3*1.4,0);

\draw[photon] (3*1.4,1*1.4)--(4*1.4,1*1.4);
\draw[photon] (3*1.4,0)--(4*1.4,0);

\node[black] at (2*1.4,0.7*1.4) {$S$};
\node[black] at (1.45*1.4,0.2*1.4) {$\widehat{D}_5$};
\node[black] at (2.4*1.4,0.2*1.4) {$\widehat{D}_5$,$\widehat{L}_5$};
\end{scope}
\end{tikzpicture}
\caption{Schematic diphoton production and decay. The solid lines in the loops stand for both scalar and fermionic components of the superfields $\hat D_5,\,\hat L_5$.}
\label{fig:diphoton_diagram}
\end{figure}

The alternative scenario, where decays of the scalar to vector-like leptons are kinematically forbidden, typically leads to more promising LHC signatures. In the absence of the superpotential term\footnote{Assuming a non-vanishing coupling $\lambda_S$ leads to non-negligible mixing of $h_s$ and $a_s$ with the MSSM Higgses. This opens up new tree-level decay modes which dominate over the loop induced decay modes. The most striking of which are decays into $W^+W^-$ and $ZZ$ if $h_s$ mixes with the standard model-like Higgs. The other possibility is mixing with the heavy Higgs. In which case the dominant decay modes are to standard model fermions.} $\lambda_S \hat S \hat H_u \hat H_d$ the decays of both the CP-even and odd component scalars, $h_s$ and $a_s$ respectively, proceed through the same loops as the production modes as shown in Fig.~\ref{fig:diphoton_diagram}. This leads to both dijet and diphoton signatures at the LHC, where the diphoton channel also receives additional enhancement due to vector-like leptons running in the aforementioned loops. Note that other diboson channels are also possible through the same loop diagrams, however the diphoton channel is experimentally more promising. 

Preliminary data of the CERN LHC \cite{Aad:2014ioa,Khachatryan:2015qba,ATLAS:2015,CMS:2015dxe} hinted at the existence of a new resonance at 750~GeV with a large diphoton cross section $\sigma_{\gamma\gamma}$ (defined as the product of production cross section and diphoton branching ratio) which at the time of writing appears to be a statistical fluctuation after the accumulation of more data \cite{CMS:2016crm,ATLAS:2016eeo}. Note, however, that cross sections of $\sigma_{\gamma\gamma}\lesssim 3~$fb are still compatible with data.   Motivated by this excess we illustrate the discovery potential in such a channel: Fig.~\ref{fig:bino_coannihilation} shows the maximum diphoton cross section at 13~TeV. This result is based on a number of assumptions. Firstly we have shown the sum of the cross sections obtained for both $h_s$ and $a_s$ as calculated using {\tt SARAH/SPheno} assuming $m_{h_s} = m_{a_s} =750$~GeV. In addition, the contribution from $h_s$ can be enhanced via large supersymmetry-breaking trilinear couplings $T_L S \tilde{\bar{L}}_5 \tilde L_5$ although this possibility is limited by vacuum stability~\cite{Nilles:2016bjl,Salvio:2016hnf}. Subsequently we have taken $T_L$ to be the maximum value compatible with the assumption of a stable vacuum.  
We stress that the results we find can be generalised to other values of the mass and production rate at the LHC without affecting the dark matter properties: in contrast to the singlino mass and couplings, the singlet mass and couplings depend on the soft-breaking parameters in the model.

\section{Conclusion}

In this letter we proposed an interpretation of the galactic center gamma ray excess in terms of thermal WIMPs. We showed that if dark matter annihilates into particles with long decay chains, the required soft gamma ray spectrum is obtained for dark matter masses as large as $500~\text{GeV}$. As a realization we considered the minimimal supersymmetric standard model extended by a singlet and a vector-like 5-plet of SU(5). The singlet fermion is identified with the dark matter and naturally obtains a relic density $\Omega h^2 \simeq 0.12$ through standard thermal freeze-out. The vector-like leptons which occur as final states in its annihilation induce a continuum of gamma rays. We find two windows in which the spectrum fits the \fermiLAT excess. The first one requires dark matter masses $100-150\:\text{GeV}$ and features electroweak gauge bosons as products of the vector lepton decays. A second window at dark matter masses $300-500\:\text{GeV}$ opens if the decay of the vector-leptons through the heavy MSSM Higgs bosons is kinematically accessible. In this case the long cascade decay of the heavy Higgses produces the required GeV gamma rays. For the heavy singlet fermion, proper normalization of the gamma ray flux requires either a rather steep dark matter profile, or a slight enhancement of the annihilation cross section today compared to the early universe. We provided simple thermal mechanisms leading to this result.

As a possible test of this scenario we suggested to search for the scalar superpartner of dark matter which could induce striking dijet or diboson signatures at the LHC. As a showcase we used the previous hints of ATLAS and CMS for a diphoton excess at 750~GeV and found that the diphoton cross section at this energy can reach values up to $\mathcal O($few fb). This result easily generalizes to other masses and suggests a high discovery potential. In addition to the scalar, the proposed model requires vector-like leptons and quarks close to the current experimental limits.

\section*{Acknowledgements}

We would like to thank Hans Peter Nilles and Matthew McCullough for interesting discussions as well as Herbi Dreiner for helpful comments on the manuscript. 
MEK is supported by the DFG Research Unit 2239 ``New Physics at the LHC'', MWW by the SFB-Transregio TR33 ``The Dark Universe".

\bibliography{HeavyWIMPS.bib}
\bibliographystyle{h-physrev}

\end{document}